# Ecosystem Evolution and Drivers across the Tibetan Plateau and Surrounding Regions

Yiran Xie[1], Xu Wang[1], Yatong Qian[2], Teng Liu[2, 4, 5], Hao Fan[2] *, Xiaosong Chen[2, 3, 6] *

[1] School of National Safety and Emergency Management, Beijing Normal University, Zhuhai, 519087, China

[2] School of Systems Science, Beijing Normal University, Beijing, 100875, China

[3] Institute for Advanced Study in Physics and School of Physics, Zhejiang University, Hangzhou 310058, China

[4] Earth System Modelling, School of Engineering and Design, Technical University of Munich, Munich, Germany

[5] Potsdam Institute for Climate Impact Research, Potsdam, Germany

[6] State Key Laboratory of Earth Surface Processes and Resource Ecology, Beijing Normal University, Beijing, 100875, China

* Corresponding authors.

E-mail addresses: fanhao_geo@163.com (H. Fan), chenxs@bnu.edu.cn (X. S. Chen).

# Abstract

The Tibetan Plateau (TP) and surrounding regions, vital to global energy and water cycles, are profoundly influenced by climate change and anthropogenic activities. Despite widespread attention to vegetation greening across the region since the 1980s, its underlying mechanisms remain poorly understood. This study employs the eigen microstates method to quantify vegetation greening dynamics using long-term remote sensing and reanalysis data. We identify two dominant modes that collectively explain more than 61% of the vegetation dynamics. The strong seasonal heterogeneity in the southern TP, primarily driven by radiation and agricultural activities, is reflected in the first mode, which accounts for 46.34% of the variance. The second mode, which explains 15% of the variance, is closely linked to deep soil moisture ($SM_3$, 28 cm to 1 m). Compared to precipitation and surface soil moisture ($SM_1$ and $SM_2$, 0 to 28 cm), our results show that deep soil moisture exerts a stronger and more immediate influence on vegetation growth, with a one-month response time. This study provides a complexity theory-based framework to quantify vegetation dynamics and underscores the critical influence of deep soil moisture on greening patterns in the TP.

# 1. Introduction

Vegetation plays a crucial role in biophysical and biogeochemical processes within ecosystems, influencing the exchange of water, energy, and carbon fluxes between terrestrial surfaces and the atmosphere (Bonan et al., 1992; Piao et al., 2020; Xu et al., 2022; Zhang et al., 2024b). In recent decades, global vegetation greening has primarily resulted from $CO_2$ fertilization, shifts in climate patterns, and alterations in land management practices (Piao et al., 2020; Li et al., 2024; Chen et al., 2019a). However, vegetation ecosystems worldwide are increasingly threatened by intensifying land use and extreme events such as heatwaves, droughts, and wildfires (Fan et al., 2023; Skinner et al. 2018; Trnka et al., 2014). Understanding how vegetation responds to climatic variability and human interventions is critical for future environmental adaptation (Zeng et al., 2017; Gong et al., 2024).

The Tibetan Plateau (TP), often referred to as the "third pole" and the "Asian water tower," provides essential ecosystem services across Eurasia, such as water regulation, climate moderation, and soil conservation (Zhang et al., 2015; Huang et al., 2023; Yao et al., 2015). Currently, the TP is experiencing rapid warming, with temperatures rising at approximately twice the global average rate over the past 50 years, potentially emerging as a tipping element in the Earth's system (Liu et al., 2023; Kuttippurath et al., 2023). Given the vulnerability and sensitivity of the ecosystem across the TP and surrounding regions (spanning 25°N-45°N, 65°E-105°E), vegetation in these regions is significantly influenced by climate change (Wang et al., 2023), which in turn affects the global climate system through alterations in the carbon cycle (Chen et al., 2019b), evapotranspiration (Forzieri et al., 2017; Shen et al., 2015) and surface albedo (Zeng et al., 2017). Investigating the long-term vegetation dynamics and underlying mechanisms across the TP and surrounding areas is crucial.

Greening trends on the TP have become increasingly prominent, driven by rising temperatures and improved water availability. Utilizing metrics such as the Normalized Difference Vegetation Index (NDVI) and Leaf Area Index (LAI), significant vegetation greening has been observed across the TP, particularly in the northeastern regions. These trends are largely attributable to factors like intensive irrigation, reduced snow cover, and enhanced precipitation (Li et al., 2020; Maina et al., 2022). Additionally, solar-induced chlorophyll fluorescence (SIF) measurements further confirm the robust greening in these areas, demonstrating the ecosystem's responsiveness to climatic and hydrological changes (Anniwaer et al., 2024). Despite the widespread greening, the net primary productivity (NPP) of vegetation across most TP regions remains relatively stable, as indicated by a low coefficient of variation (Zhang et al., 2021b). Advances in satellite remote sensing and reanalysis datasets have been instrumental in uncovering these large-scale vegetation dynamics, characterized by significant spatial heterogeneity and complexity. These insights collectively enhance our understanding of the greening processes on the Tibetan Plateau and their interactions with ongoing climate change.

Previous research has predominantly focused on pixel or regional scale analyses, often overlooking the intricate interactions within vegetation ecosystems. This study explores the evolutionary patterns of vegetation dynamics within a systemic framework.

The eigen microstates method, derived from statistical physics, has been effectively applied to analyze emergent phenomena and evolutionary dynamics within non-equilibrium complex systems, including those in economics, climate, and ecology (Hu et al., 2019, 2023; Liu et al., 2022; Sun et al., 2021; Wang et al., 2024). This approach divides complex systems with multiple components and high-dimensional nonlinear interactions into several mutually independent eigen microstates. By examining the spatial distribution and temporal evolution of each eigen microstate, systemic changes and their underlying mechanisms can be elucidated. In addition to the eigen microstates analysis, we assess vegetation stability by examining spatiotemporal irregularities, providing additional insights into regional vegetation dynamics.

## 2. Data and methods

**2.1 Study area**

Our study focuses on the TP and surrounding regions (25°N–45°N, 65°E–105°E; Fig. 1a), characterized by diverse topography with an average elevation exceeding 2 km. The TP, as the world's highest plateau, exceeds 4 km in elevation (Fig. 1b). The climate varies significantly, with annual precipitation ranging from over 800 mm in the southeast to less than 200 mm in the northwest, while temperatures range from below -10°C to over 20°C. These climatic and topographic variations give rise to diverse vegetation patterns (Fig. 1c), with the TP being primarily covered by grasslands, meadows, and shrubs. The southeastern TP, influenced by the South Asian summer monsoon, supports forests and shrublands (Wang et al., 2017b), while the northwestern TP, shaped by westerlies, is dominated by deserts and grasslands. In contrast, the Indus and Ganges River basins are primarily croplands (Kuttippurath and Kashyap, 2023).

**2.2 NDVI and meteorological data**

NDVI, derived from red and near-infrared reflectance, is a validated indicator of vegetation activity and is widely used to assess vegetation dynamics (Anniwaer et al., 2024; Li et al., 2020; Tucker et al., 2005). We used the third-generation Global Inventory Monitoring and Modelling System (GIMMS) NDVI dataset from the Advanced Very High Resolution Radiometer (AVHRR) instruments (Holben, 1986; Pinzon and Tucker, 2014). It has a spatial resolution of 1/12° and a temporal resolution of 15 days, spanning from 1982 to 2015. To reduce the effects of clouds, sandstorms, and haze, we applied the 15-day maximum value composite (MVC) method to generate monthly NDVI datasets. Pixels with an average NDVI below 0.1 during 1982-2015 were excluded to minimize the influence of bare land and sparse vegetation.

This study analyzed the influence of monthly surface air temperature (T2m), total precipitation (Pp), soil moisture across three layers ($SM_1$: 0-7 cm, $SM_2$: 7-28 cm, $SM_3$: 28-100 cm), and surface solar radiation (SSR) on vegetation growth from 1982 to 2015. In this study, $SM_1$ and $SM_2$ are classified as surface soil moisture, while $SM_3$ is categorized as deep soil moisture. Unless otherwise specified, soil moisture refers to all three layers collectively. These variables are from the ERA5-Land dataset with a spatial resolution of 0.1°, while total cloud cover is from ERA5 with a spatial resolution of 0.25°. To match the spatial resolution of NDVI, all meteorological data were resampled to 1/12° using the nearest neighbor resampling method.

**2.3 Eigen microstates method**

In an ecosystem comprising $N$ grids, the NDVI for grid $i$ at time $t$ is defined as $S_i(t)$. The fluctuation of NDVI of grid $i$ at time $t$ is defined as $\delta S_i(t) = S_i(t) - \langle S_i(t) \rangle$, where $\langle S_i(t) \rangle = \frac{1}{T}\sum_{t=1}^{T} S_i(t)$ represents the average NDVI during the period $T$. Due to the strong eco-geographical diversity in the study area, vegetation cover shows significant variations and variance differences. A preferable approach to characterize NDVI fluctuation is

$$\delta S_i(t) = \frac{S_i(t) - \langle S_i(t) \rangle}{\sqrt{\frac{1}{T}\sum_{t=1}^{T}\left[S_i(t) - \langle S_i(t) \rangle\right]^2}}. \tag{1}$$

A microstate of a system is delineated by the collective fluctuations of $N$ grids at time $t$. Therefore, a statistical ensemble for the complex system involving $T$ microstates can be formulated into an $N \times T$ matrix $\boldsymbol{A}$ with the element $A_{it} = \delta S_i(t)/\sqrt{C_0}$, where $C_0 = \sum_{t=1}^{T}\sum_{i=1}^{N} \delta S_i^2(t)$ is the normalization factor.

According to the singular value decomposition, the ensemble matrix $A$ can be factorized as

$$A = U \cdot \Sigma \cdot V^T = \sum_{I=1}^{r} \sigma_I A_I^e = \sum_{I=1}^{r} \sigma_I U_I \otimes V_I, \qquad (2)$$

Where $\Sigma$ is an $N \times T$ diagonal matrix with diagonal elements $\sigma_I$ satisfying $\sum_{I=1}^{r} \sigma_I^2 = 1$. $A_I^e = U_I \otimes V_I$ is an $N \times T$ matrix with the element $\left(A_I^e\right)_{it} = U_{iI} V_{tI}$. Therefore, the statistical ensemble of the system can be determined through a linear superposition of eigen ensemble $A_I^e$ with the probability of $\sigma_I^2$. From the statistical ensemble, we can obtain not only the eigen microstate $U_I$, but also its temporal evolution $V_I$. A detailed description of the eigen microstates method is provided in Text S1.

Because of nonlinear interactions and feedback loops between and within vegetation, climate, hydrology, topography, and human activities in the TP and surrounding regions, vegetation dynamics exhibit significant complexity, characterized by emergent collective behaviors and pronounced spatiotemporal heterogeneity (Yao et al., 2015; Anniwaer et al., 2024). Solely relying on traditional statistical methods, such as regression and correlation analysis, is insufficient to effectively capture these complexities. The eigen microstate method, derived from statistical physics, provides a more physically meaningful perspective to study the systematic evolution in vegetation dynamics by extracting dominant patterns and uncovering their underlying drivers (Wang et al., 2024). This approach is well-suited for analyzing emergent phenomena within non-equilibrium complex systems (Ma et al., 2024; Sun et al., 2021) while accommodating vegetation variations across different temporal scales, from long-term trends to seasonal and interannual fluctuations.

## 2.4 Stability analysis

For a given time series $X_t$, the randomness and stationarity can be measured by the autocorrelation function, defined as

$$C(\tau) = \frac{\text{Cov}(X_t, X_{t+\tau})}{\sqrt{\text{Var}(X_t)\text{Var}(X_{t+\tau})}} = \frac{E\left[(X_t - E(X_t))(X_{t+\tau} - E(X_{t+\tau}))\right]}{\sqrt{E\left[(X_t - E(X_t))^2\right] \cdot E\left[(X_{t+\tau} - E(X_{t+\tau}))^2\right]}}, \quad (3)$$

where the time lag $\tau$ belongs to [-$\tau_{max}$, $\tau_{max}$]. However, due to noise superimposed on the collected data and underlying trends of unknown origin, a direct calculation of $C(\tau)$ is usually not appropriate. Here, we introduce an advanced weighted autocorrelation function $W_{ACF}$ proposed by Meng et al. (2023) to quantify memory strength:

$$W_{ACF} = \frac{\max(|C(\tau)|) - \text{mean}(|C(\tau)|)}{\sqrt{\text{Var}(|C(\tau)|)}} \equiv \frac{1 - \text{mean}(|C(\tau)|)}{\sqrt{\text{Var}(|C(\tau)|)}}, \quad (4)$$

where $\max(|C(\tau)|)$ and $\text{mean}(|C(\tau)|)$ represent the maximum and mean values of the absolute autocorrelation function $|C(\tau)|$ respectively. Considering the seasonality of vegetation growth, $\tau_{max} = 4$ months is selected for the primary analysis in this study. We also calculated $\tau_{max}$ values of 3, 5, and 6 months, and the results were consistent. We assess vegetation stability and stochasticity by applying $W_{ACF}$ to NDVI, where higher values indicate greater disorder and instability. Meng et al. (2023) introduce an advanced weighted power spectral density ($W_{PS}$) based on Welch's method (Welch, 1967) to analyze frequency characteristics (see Text S2 for details).

## 2.5 Trend analysis

The Linear Least Squares Regression method was utilized to assess the trends of NDVI and meteorological factors at both pixel and regional scales. The trend is defined as the slope of the regression line:

$$Slope = \frac{\sum_{i=1}^{T} y_i t_i - \frac{1}{T}\left(\sum_{i=1}^{T} y_i\right)\left(\sum_{i=1}^{T} t_i\right)}{\left(\sum_{i=1}^{T} t_i^2\right) - \frac{1}{T}\left(\sum_{i=1}^{T} t_i\right)^2}, \tag{5}$$

where *Slope* represents the trend; $T$ is the cumulative number of months during the study period, and $y_i$ is the value of NDVI or meteorological factors at time $t_i$. The positive (negative) values of the *Slope* represent an upward (downward) trend of NDVI or climate variables. We calculated 95% confidence intervals and verified trend significance using a *t*-test ($p < 0.05$).

## 3. Results

### 3.1 Spatiotemporal variation of vegetation across the TP and surrounding regions

Spatiotemporal variation patterns of NDVI across the TP and surrounding regions are presented in Fig. 2. From 1982 to 2015, the annual average NDVI shows a significant upward trend ($y = 0.00052x + 0.33$, $p < 0.01$, $R^2 = 0.53$), indicating persistent vegetation greening (Fig. 2a). The mean NDVI during this period was 0.34. Seasonal NDVI variations are depicted in Fig. S1a with a peak of 0.42 in summer (August) and a trough of 0.29 in winter (December). Across all four seasons, the annual mean NDVI exhibits significant increasing trends with distinct internal variations (Fig. S2a, d, g, j). Specifically, summer maintains the highest values, while spring and autumn show faster growth rates, and winter displays moderate variation. The spatial distribution of average NDVI exhibits notable heterogeneity, with higher values observed in the northeastern and southeastern margins of the TP and lower values in the northwest (Fig. 2b). NDVI increases from west to east (Fig. S1b) and decreases from south to north (Fig. 2c), with averages of 0.39 (25°N - 35°N) and 0.22 (35°N - 45°N), respectively. The eastern region, including the southeastern part of the plateau and the northern Indo-China Peninsula, is characterized by a warmer and more humid climate, providing more favorable conditions for vegetation growth (Li et al., 2020). In contrast, the western region, although it includes the dense vegetation of the Indo-Gangetic Plain, also encompasses the high-altitude Tibetan Plateau and the arid to semi-arid zones of Central Asia, where harsher climates and lower precipitation lead to lower NDVI values (Li et al., 2021b). Seasonal spatial patterns align with the annual distribution (Fig. S2b, e, h, k).

Grasslands in the eastern and northwestern regions exhibit distinct seasonal changes, with higher NDVI in summer and autumn, whereas agricultural and forested areas in the southern low-latitudes exhibit minimal seasonal variation.

In terms of the NDVI trend, approximately 69% of regions exhibit significant increases, primarily in northern India and eastern TP in China (Fig. 2d, e). Conversely, 31% of regions exhibit a declining trend, with 16% showing a significant decrease, primarily located between 40°N and 45°N (Fig. 2f). Significant vegetative degradation has been observed in southern Kazakhstan, with northeastern India also attracting considerable attention (Yang et al., 2022; Li et al., 2021b). Seasonally, the overall trend shows an upward trajectory, particularly in spring and autumn (Fig. S3).

**3.2 Stability of vegetation variation across the TP and surrounding regions**

In systems characterized by highly nonlinear interactions, rapid changes in control parameters can profoundly affect system stability (Liu et al., 2023). Given the marked temperature increases observed in the TP region, it is imperative to examine the stability of vegetation dynamics within the context of ongoing greening trends. Standard deviation (SD), which quantifies data dispersion, is commonly used to assess value variability. Overall, SD increases with longitude, but no distinct trend is observed along latitude (Fig. 3c, d). The spatial pattern of SD for NDVI generally corresponds to its spatial mean distribution (Fig. 3a). Notably, regions with higher SD are primarily found in the northeastern TP and around the Tianshan Mountains, where grasslands experience significant seasonal changes, resulting in greater data dispersion. Conversely, the southeastern forests exhibit minimal seasonal fluctuations, leading to lower SD.

Since the SD does not effectively capture the irregularity of vegetation changes, as demonstrated by its invariance when the data is randomly shuffled. We further calculate $W_{ACF}$ and $W_{PS}$ to assess the memory strength and the frequency characteristics. The spatial patterns of $W_{ACF}$ and $W_{PS}$ are highly similar, and their temporal evolution demonstrates stable oscillations and synchrony, indicating consistent irregularity and frequency variability (Fig. 3b, Fig. S4c, d). To further test the robustness of the results, we varied $\tau_{max}$ values to 3, 5, and 6 months and observed consistent spatiotemporal patterns between $W_{ACF}$ and $W_{PS}$ (Fig. S4). Therefore, the subsequent analysis focuses on $W_{ACF}$ variations.

As shown in Fig. 3b, higher $W_{ACF}$ values are observed in the southern regions indicating greater irregularity and more rapid variations. Additionally, $W_{ACF}$ decreases with increasing latitude and longitude (Fig. 3d). The region between 25°N-35°N and 65°E-75°E exhibits higher $W_{ACF}$ values and lower SD, motivating further investigation into the drivers of rapid and irregular vegetation changes in this region. Specifically, we calculate the correlation between $W_{ACF}$ and critical environmental variables influencing vegetation growth, including T2m, Pp and three-layer soil moisture (Fig. 3e, f, Fig. S5). Results reveal a positive correlation between T2m and $W_{ACF}$ ($R = 0.43$) and negative correlations with Pp ($R = -0.55$) and soil moisture ($SM_1$-$SM_3$: $R = -0.59, -0.68, -0.66$, respectively). Moreover, soil moisture at three layers shows a stronger negative correlation than Pp. These findings suggest that warming and reduced water availability, especially in the soil, may contribute to a decline in vegetation stability.

## 3.3 Key drivers of vegetation greening

The first two sections delineate the spatiotemporal patterns of vegetation dynamics, assessing nonlinear fluctuations and stability at both pixel and regional scales, providing valuable insights into the complexity of the study area. Building on this, we apply the eigen microstates method to investigate the systematic evolution of vegetation and identify its key drivers, considering the interactions within the ecosystem.

### 3.3.1 Effects of radiation and agricultural activities on seasonal cycle heterogeneity

The two largest eigen microstates of NDVI explain 61% of the vegetation dynamics, with the first mode accounting for 46.34%. Temporal evolution $V_1$ (Fig. 4c) exhibits pronounced interannual variability, characterized by a significant annual peak in its spectrum analysis (Fig. 4e). Additionally, $V_1$ remains positive from May to September, peaking in August, while it turns negative from December to April, reaching its lowest point in February (Fig. 5a). Spatially, $U_1$ (Fig. 4a) displays positive signals in 80% of the area, with negative signals in 20% of the regions, including the Indo-Gangetic Basin (region 1), central-eastern Himalayas (region 2), and northern Myanmar (region 3). These results demonstrate that the first eigen microstate effectively captures the seasonal vegetation cycle, with higher NDVI values observed from May to September and lower values from October to April, aligning with established growing season definitions (Liu et al., 2024; Zhang et al., 2023a; Tang et al., 2023).

However, significant heterogeneity in the seasonal cycle is observed in the southern region, primarily driven by agricultural activities and radiation (Fig. 4a).

Specifically, Region 1, as one of the major agricultural areas in South Asia, experiences intense agricultural activities, characterized by three distinct cropping seasons: Kharif (June-September), Rabi (October-February), and Zaid (March-May) (Sharma et al., 2024; Kuttippurath and Kashyap, 2023). The Kharif season benefits from ample precipitation brought by the South Asian summer monsoon, while the Rabi season occurs post-monsoon (Wang et al., 2022; Revadekar et al., 2012). The Zaid season, a transitional period with warm and dry conditions, depends on agricultural irrigation for short-duration crops. The widespread practice of multiple cropping and crop rotation enhances land use efficiency and ensures continuous crop production across seasons (Han et al., 2022; Kumawat et al., 2023). Regions 2 and 3 are influenced by the South Asian monsoon, resulting in distinct wet and dry seasons. As shown in Fig. 6b, the summer monsoon (May-September) brings abundant rainfall and increases cloud cover, which reduces shortwave solar radiation and may lead to lower NDVI values (Yang et al., 2018; Saleska et al., 2016; Zhang et al., 2016; Nemani et al., 2003). We further computed the spatial correlation between the temporal evolution $V_1$ of NDVI and SSR (Fig. 6a). The significant negative correlations are observed in most areas of regions 2 and 3, closely resembling the pattern shown in Fig. 4a, providing additional support for the role of radiation in driving the seasonal heterogeneity in these regions.

**3.3.2 The critical role of soil moisture in vegetation growth**

The temporal evolution of the second eigen microstate, which accounts for 15.00% of the total variance, exhibits an annual periodic signal (Fig. 4d, f). It peaks in September, coinciding with the end of the South Asian summer monsoon, and reaches its lowest value in May during the spring warm period (Fig. 5b). Spatially, a negative signal is observed in the northwestern region, which is characterized by high temperatures and strong evapotranspiration in summer, leading to drier conditions. However, it receives more moisture in spring and winter from the Westerlies, snowmelt, and regional precipitation. A strong positive signal is found in northern India, northern Myanmar, and the southern Hengduan Mountain region, where abundant water resources are available during the South Asian summer monsoon. These findings highlight a close association between the second eigen microstate and water.

To further explore the underlying significance of the second eigen microstate, we examined its relationships with water-related environmental variables, including three-layer soil moisture and Pp. We derived their spatial patterns and temporal evolutions using the eigen microstates method. As shown in Fig. 7, $U_1$ of the Pp and three-layer soil moisture exhibit strong correlations with $U_2$ of NDVI, with correlation coefficients of 0.72, 0.82, 0.81, and 0.74 ($p < 0.01$). Power spectrum analysis reveals a significant annual cycle in $V_1$ of meteorological factor, consistent with the findings in $V_2$ of NDVI (Fig. S6). However, the correlation between $V_1$ of Pp and $V_2$ of NDVI is weak and does not pass the significance test. As soil depth increases, the correlation strengthens from 0.22 to 0.35 and 0.59 ($p < 0.01$), highlighting the critical influence of deeper soil moisture on vegetation growth.

Additionally, a lag effect between NDVI and both Pp and soil moisture is observed in Fig. 8. The correlation improves significantly when Pp lags by 1-2 months ($R = 0.5$, 0.84, $p < 0.01$). Surface soil moisture exhibits peak correlations at a 2-month lag ($R = $ 0.69 and 0.78), while deep soil moisture shows the strongest correlation with NDVI at a 1-month lag ($R = 0.79$), indicating a rapid vegetation response to the deep soil moisture.

# 4. Discussion

## 4.1 Widespread vegetation greening across the TP and surrounding regions

Vegetation greenness has been increasing globally since at least 1981, particularly in high-latitude regions, agricultural areas, and afforestation zones (Piao et al., 2020). We also observe an overall improving trend in vegetation dynamics across the TP and surrounding regions from 1982 to 2015, particularly notable in India and China (Wang et al., 2017a; Kuttippurath and Kashyap, 2023).

However, most studies have primarily focused on individual-level trends, overlooking the interactions between them. To address this gap, we applied the eigen microstates method to decompose the deseasonalized NDVI data (Text S3). As shown in Fig. 9, the first mode reveals a prominent greening trend across the study area (Zhang et al., 2024a; Wu et al., 2021). $U_1$ of the deseasonalized NDVI exhibits positive signals across most of the region, with notable negative signals near the Tianshan Mountains and central Asia. $U_1$ exhibits a strong correlation with the trend pattern shown in Fig. 2e ($R = 0.82$, $p < 0.01$). Furthermore, $V_1$ demonstrates a clear upward trend, consistent with the long-term trend of deseasonalized NDVI from 1982 to 2015, showing a correlation coefficient of 0.73 ($p < 0.01$). These findings not only provide stronger evidence for the greening trend in the study area but also highlight the robustness of the eigen microstates method in understanding vegetation dynamics.

## 4.2 Deep soil moisture as a crucial driver of vegetation greening

We further observed a significant upward trend in $V_2$ of NDVI over the years,

highlighting its substantial contribution to the overall vegetation greening (Fig. 10a). To identify the drivers of greening, we calculated the correlations between the first mode of Pp and soil moisture at all three layers with the second mode of NDVI (Fig. 10b). Soil moisture exhibited stronger correlations with NDVI that increased with soil depth ($R$ = 0.52, 0.61, 0.72, $p$ < 0.01), in contrast to the weak correlation between Pp ($R$ = 0.30, $p$ < 0.1). Moreover, the first mode of deep soil moisture showed a significant upward trend, underscoring its key contribution to vegetation greening compared to Pp and surface soil moisture.

Soil moisture directly regulates water availability to plants, exerting a profound influence on their growth patterns (Wang et al., 2017a, Wang et al., 2025). Insufficient soil moisture inhibits photosynthesis and biomass accumulation in plants, potentially leading to mortality, while adequate moisture promotes root development and canopy growth (Xue and Wu, 2024; Zhang et al., 2023b; Li et al., 2022; Yang et al., 2022). Deep soil moisture offers a stable, long-term water supply to vegetation, mitigating extreme drought impacts due to its low evaporation and weak sensitivity to short-term precipitation variability (Fan et al., 2017; Li et al., 2021a). Variations in soil moisture in the root zone impact crop yields and vegetation carbon fixation efficiency (Green et al., 2019; Xue and Wu, 2024).

Chen et al. (2019a) studied global vegetation leaf area changes from 2000 to 2017, highlighting the significant contributions of China and India to global greening and identifying human land-use practices, such as agriculture and afforestation, as key driving factors. Kuttippurath and Kashyap (2023) further analyzed vegetation changes in India from 2000 to 2019, finding that croplands drive (86.5% contribution) the greening of India. The enhanced greening of croplands was primarily driven by improved irrigation facilities and effective agricultural management practices, such as mechanization and fertilization. These studies highlight the direct impact of human activities on vegetation greening, while our research further reveals the critical role of soil moisture in vegetation dynamics. For example, in the Indo-Gangetic Plain, agricultural irrigation significantly increases soil moisture, thereby promoting vegetation growth (Ambika and Mishra, 2019; Bhimala et al., 2020; Chen et al., 2024; Dubey and Ghosh, 2023; Kuttippurath and Kashyap, 2023). However, in our study region, soil moisture is influenced not only by human activities but also by the combined effects of temperature, atmospheric circulation, precipitation, snowmelt, and topography (Wang et al., 2025). This indicates that soil moisture serves as a key intermediary variable at the interface of natural and anthropogenic influences, playing a complex regulatory role in vegetation dynamics. Eigen microstates analysis reveals a significant greening trend in the TP and South Asia, linked to increasing soil moisture, while vegetation browning in Central Asia and the Tianshan Mountains may be driven by soil moisture decline (Li et al., 2020; Li et al., 2021b; Zhang et al., 2021a). Our findings reveal the critical role of soil moisture in vegetation greening within the framework of ecosystem evolution, offering new insights into vegetation dynamics.

**4.3 Uncertainties and limitations**

Our study involves some uncertainties and limitations. Resampling soil moisture data to match the spatial resolution of NDVI may result in the loss of some intricate details and geographical complexities. Additionally, NDVI derived from satellite observations could be affected by clouds and aerosol contaminations, especially in southeastern TP. Therefore, future research should integrate field observations with high-resolution remote sensing data to more accurately capture vegetation dynamics. While our study highlights the critical role of soil moisture in vegetation dynamics and reveals a significant increase in deep soil moisture, the relationship between soil moisture and vegetation change is a complex coupled process. Future research should employ multi-scale and multi-variable analytical approaches to further disentangle the interactions among soil moisture, climate variability, and human activities.

# 5. Conclusion

This study investigates the systemic evolution of vegetation across the TP and surrounding regions using long-term remote sensing and reanalysis data. The analysis of vegetation stability reveals rapid and irregular changes in the southern region, which exhibit positive correlations with T2m and negative correlations with Pp and soil moisture. We further apply the eigen microstates method to identify two dominant modes that collectively explain over 60% of vegetation dynamics. The first mode highlights significant heterogeneity in the seasonal cycle of vegetation, particularly in the southern TP, primarily driven by radiation and agricultural activities. The second mode shows that deep soil moisture exerts a stronger and more immediate influence on vegetation growth, contributing more substantially to vegetation greening than Pp or surface soil moisture. Additionally, after removing seasonal variability, the first mode further supports the observed greening trend in the study area. This study demonstrates the effectiveness of the eigen microstates method in analyzing vegetation dynamics, offering a novel insight into the significant contribution of soil moisture in vegetation greening.

## Declaration of competing interest

All authors declare that they have no conflicts of interest.

## Data availability

The AVHRR GIMMS NDVI3g dataset is available at A Big Earth Data Platform for Three Poles (https://poles.tpdc.ac.cn/en/data/9775f2b4-7370-4e5e-a537-3482c9a83d88/). The ERA5-Land datasets are available at https://cds.climate.copernicus.eu/datasets. The altitude data provided by Amatulli et al. (2018) with the resolution of 1km × 1km can be downloaded from http://www.earthenv.org/topography. The land cover data are extracted from the Climate Change Initiative Land Cover datasets of the European Space Agency (2015) at a spatial resolution of 300m × 300m (ESA/CCI viewer (ucl.ac.be)). All figures were created using Matplotlib version 3.7.2, available under the Matplotlib license at https://matplotlib.org/. Moreover, statistical analyses were conducted on the Python platform (https://www.python.org/).

## Acknowledgements

This study was supported by the National Key R&D Program of China (grant 2023YFE0109000), the National Natural Science Foundation of China (grants 12135003 and 42205178), the State Key Laboratory of Earth Surface Processes and Resource Ecology (2024-TS-05), the Fundamental Research Funds for the Central Universities (2243100009). This is ClimTip contribution 38; the ClimTip project has received funding from the European Union's Horizon Europe research and innovation programme under grant agreement No. 101137601. As Associated Partner the BNU has received funding from the Chinese Ministry for Science and Technology (MOST). The

# Figures

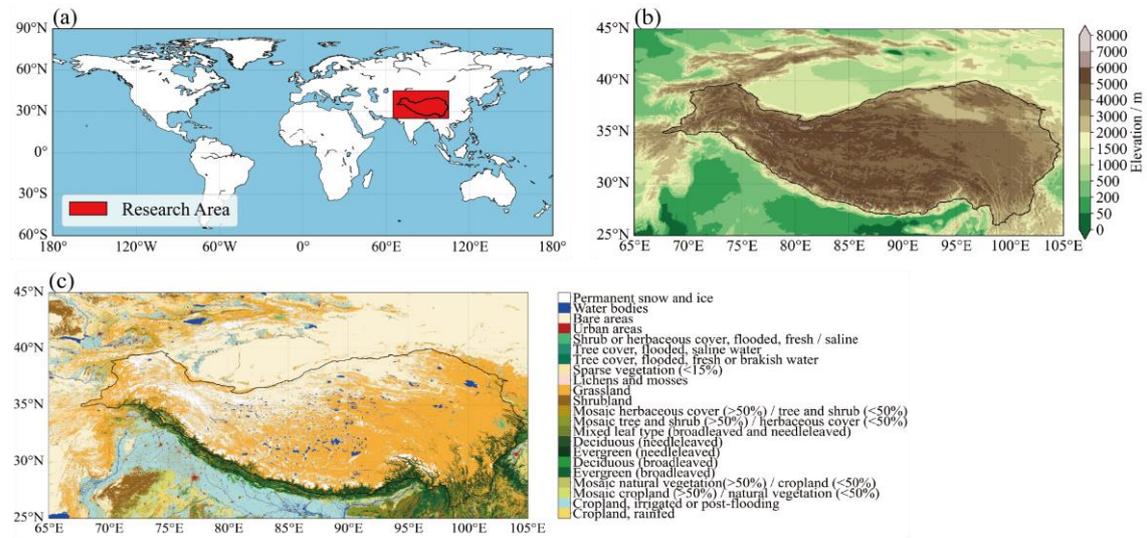

**Fig. 1. Overview of the TP and surrounding regions.** (a) Geographic location of the study area in a global context. (b) Spatial distribution of altitudes. (c) Land cover types across the study area.

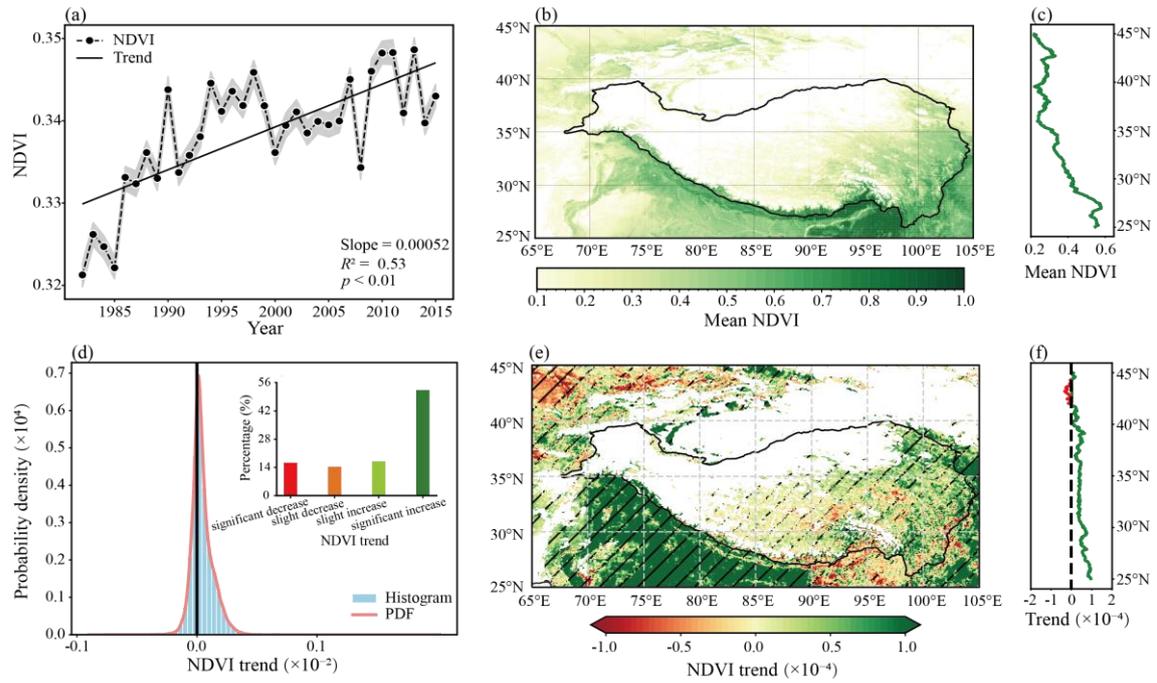

**Fig. 2. Spatiotemporal dynamics of vegetation across the TP and surrounding regions.** (a) Temporal evolution of annual average NDVI from 1982 to 2015 with the shading representing the 95% confidence interval. (b) Spatial distribution of average NDVI. (c) Average NDVI along latitude. (d) Probability density distribution of NDVI trends. (e) Spatial patterns of NDVI trends from 1982 to 2015. Shaded areas with diagonal lines indicate regions where the significance level is below 0.05. (f) Averaged trend along latitude, with red indicating negative values (declining NDVI) and green indicating positive values (increasing NDVI).

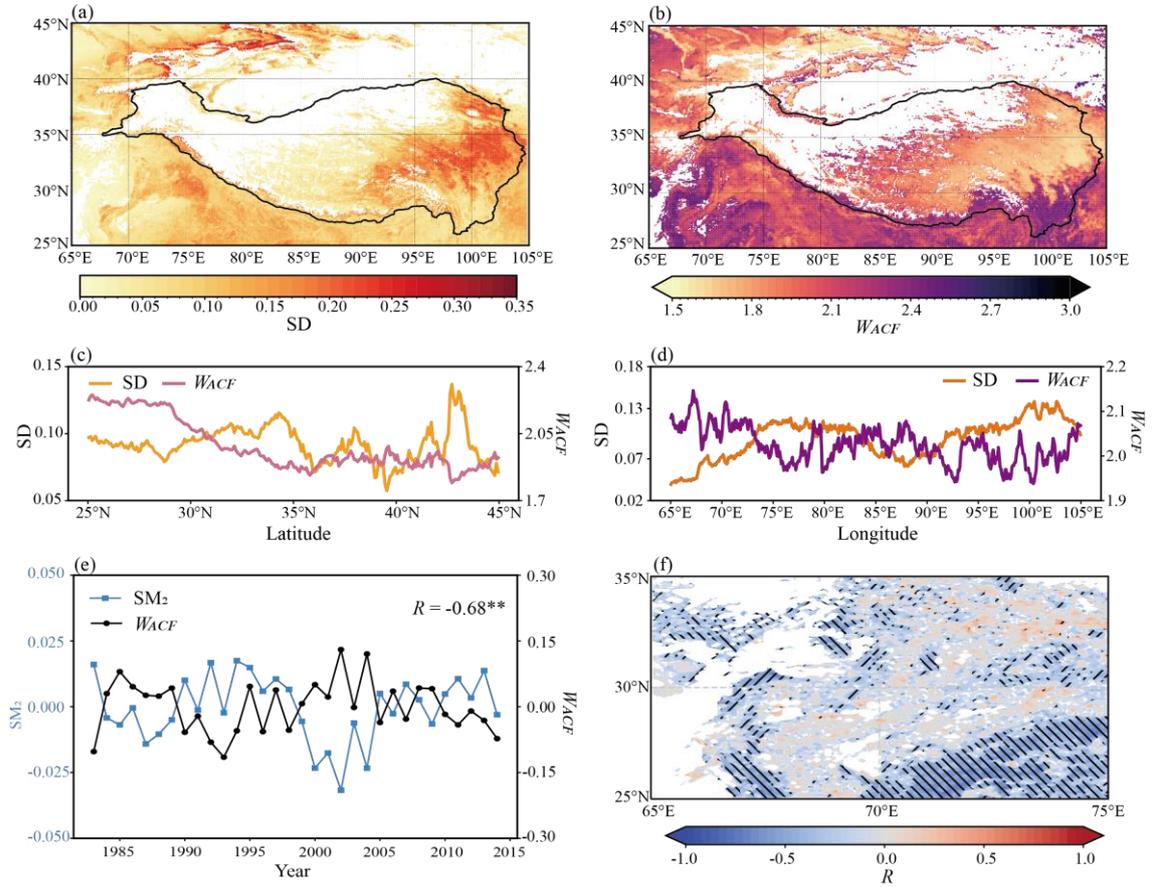

**Fig. 3. Stability of vegetation variation across the TP and surrounding regions.** Spatial distribution of (a) SD and (b) $W_{ACF}$. The (c) zonal and (d) meridional averages of SD and $W_{ACF}$. (e) Temporal variations of the annual mean $SM_2$ and $W_{ACF}$ in the region spanning 25°N-35°N, 65°E-75°E, along with their correlation coefficient ($R$). Additionally, ** indicates a significant correlation ($p < 0.01$). (f) Spatial distribution of correlation coefficient ($R$) between $W_{ACF}$ and the annual average soil moisture in the second layer. Shaded areas with diagonal lines indicate regions where the significance level is below 0.05.

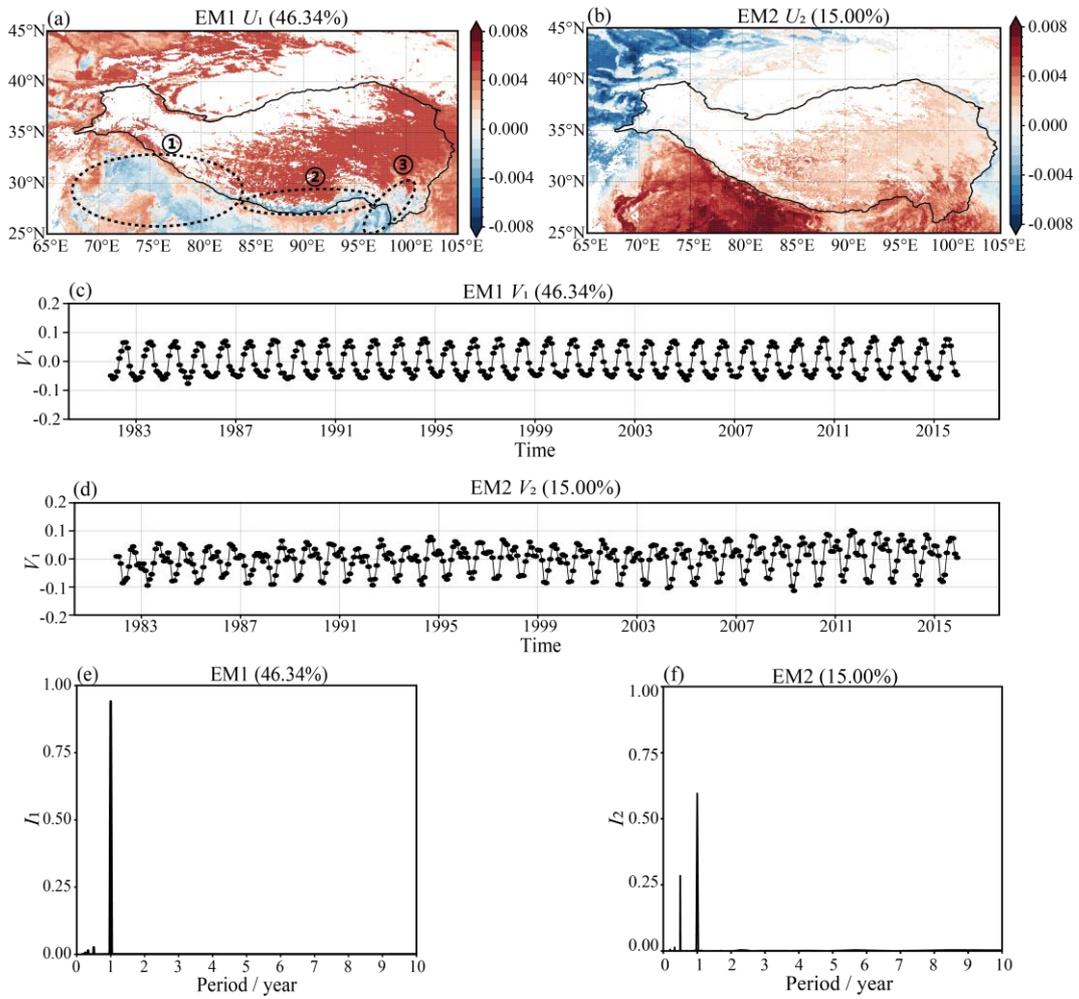

**Fig. 4. Spatial patterns and temporal evolutions of two leading modes of NDVI.** (a), (b) Spatial patterns and (c), (d) temporal evolutions of two leading modes. (e), (f) Power spectrum density of temporal evolutions of two leading modes.

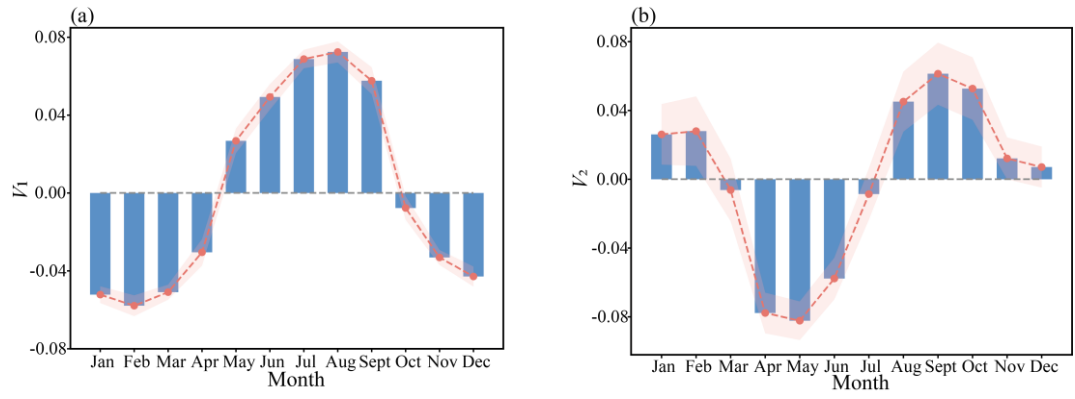

**Fig. 5. Seasonal cycle of temporal evolution $V_1$ and $V_2$ for the two leading modes of NDVI.**

Shading represents ±1 SD.

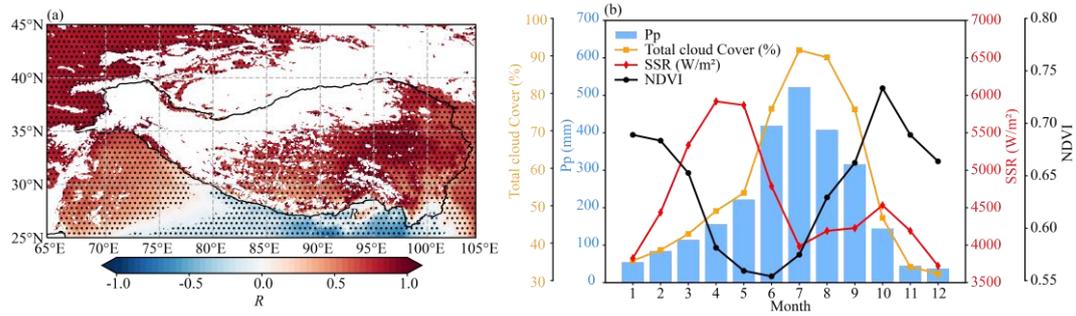

**Fig. 6. Spatial correlations and monthly temporal variations of NDVI and climate factors.** (a) Spatial correlation between temporal evolution $V_1$ of the first mode for NDVI and the original monthly SSR data. (b) Monthly variations of SSR, Pp, total cloud cover, and NDVI in grids with negative signals in region 2 and region 3.

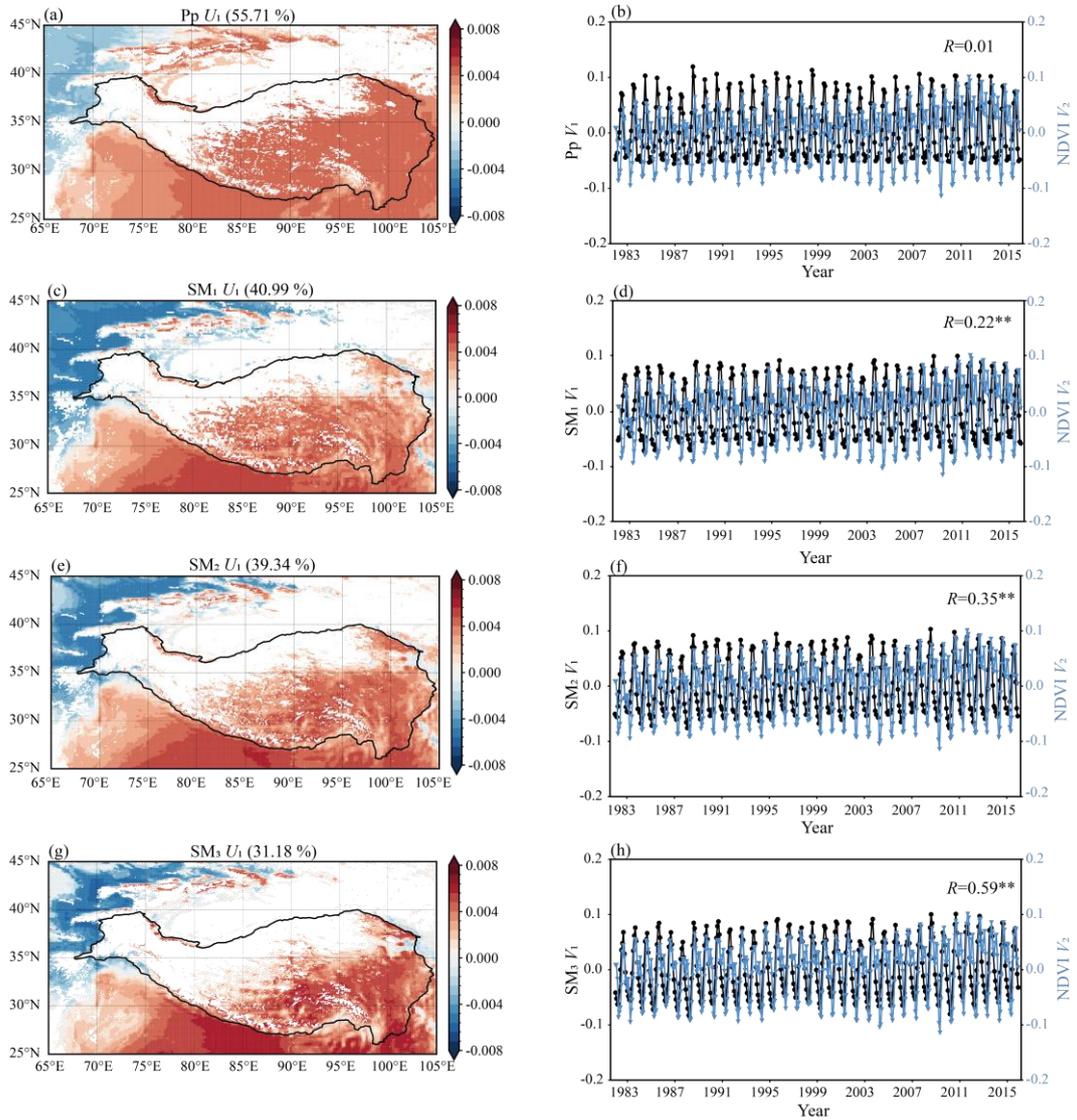

**Fig. 7. Spatial patterns and temporal evolutions of the first mode for climate variables.** (a), (c), (e), (f) Spatial patterns $U_1$ and (b), (d), (f), (g) temporal evolutions $V_1$ of the first mode for Pp, $SM_1$, $SM_2$, and $SM_3$, and temporal evolution $V_2$ of the second mode for NDVI. Correlation coefficients $R$ are provided, with ** indicating statistical significance ($p < 0.01$) and values without asterisks indicating $p > 0.1$.

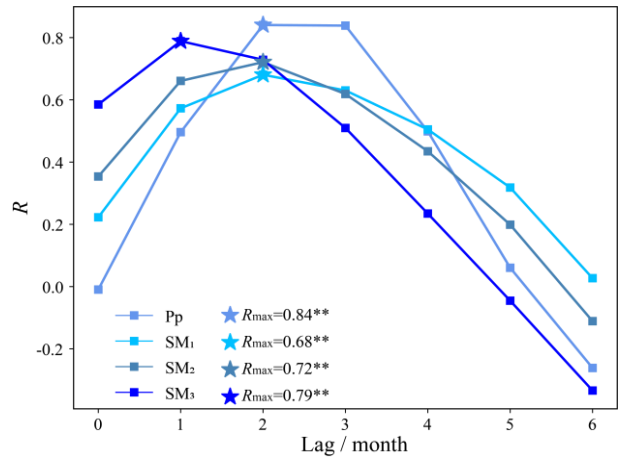

**Fig. 8. Lagged correlation coefficients between temporal evolution $V_2$ of the second mode for NDVI and temporal evolution $V_1$ of the first mode for Pp, $SM_1$, $SM_2$, and $SM_3$.** $R_{max}$ indicates the maximum correlation values. ** denotes significant correlation ($p < 0.01$).

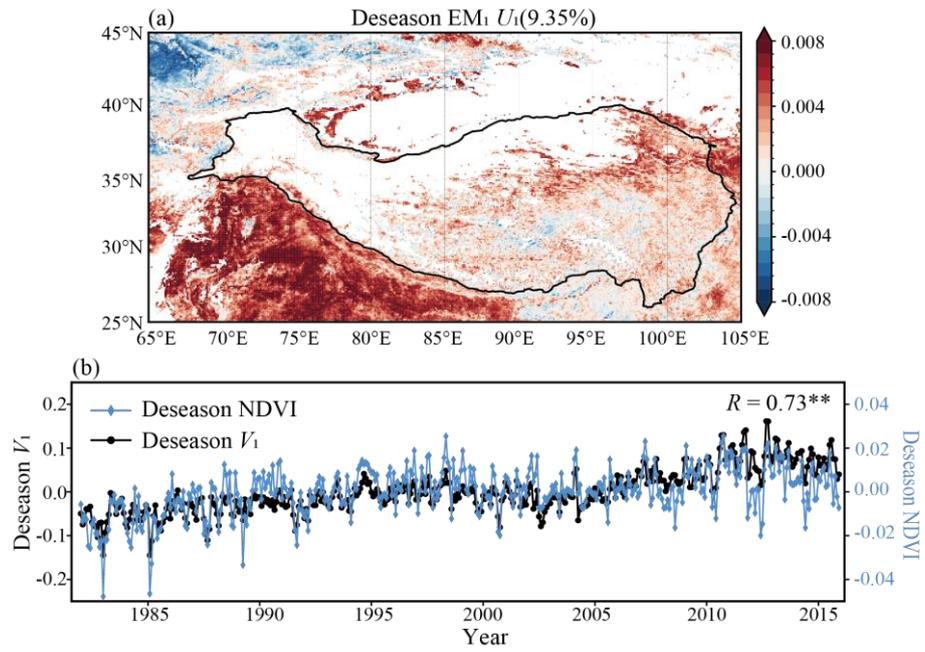

**Fig. 9. Spatial patterns and temporal evolutions of the first mode for deseasonalized NDVI.** (a) Spatial pattern of the first mode. (b) Temporal evolution of the first mode (black line) and monthly variation of deseasonalized NDVI (blue line). *R* represents their correlation coefficient, and ** indicates a significant correlation ($p < 0.01$)

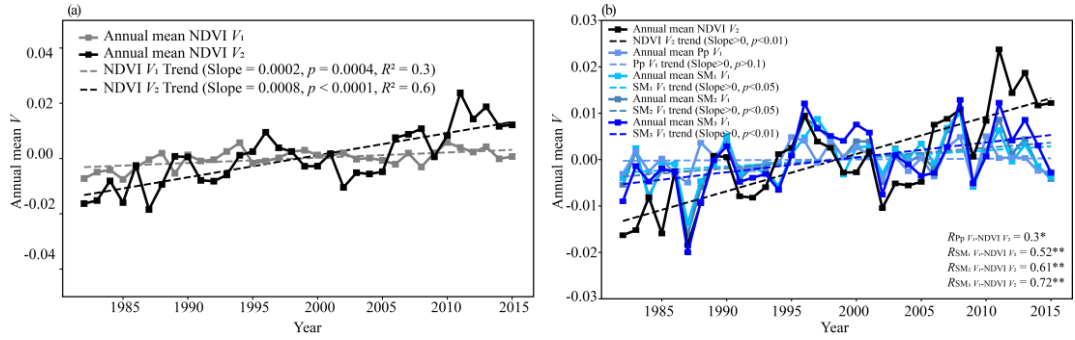

**Fig. 10. Annual mean variations and trends of the temporal evolutions of modes for NDVI and climate variables.** (a) Annual mean variations and trends of temporal evolutions for the first two modes of NDVI. (b) Annual mean variations and trends of temporal evolutions for the first mode of Pp, $SM_1$, $SM_2$, and $SM_3$ and the second mode of NDVI. $R$ represents their correlation coefficients. ** indicates statistical significance at $p < 0.01$, and * at $p < 0.05$.

**Text S1. Eigen microstates method**

In an ecosystem comprising $N$ grids, the NDVI for a grid $i$ at time $t$ is defined as $S_i(t)$. The fluctuation of NDVI of grid $i$ at time $t$ is defined as $\delta S_i(t) = S_i(t) - \langle S_i(t) \rangle$, where $\langle S_i(t) \rangle = \frac{1}{T}\sum_{t=1}^{T} S_i(t)$ represents the average NDVI during the period $T$. Due to the pronounced ecogeographical heterogeneity across the study area, vegetation cover exhibits substantial variations, leading to significant differences in variance. A preferable approach to characterize NDVI fluctuation is

$$\delta S_i(t) = \frac{S_i(t) - \langle S_i(t) \rangle}{\sqrt{\frac{1}{T}\sum_{t=1}^{T}\left[S_i(t) - \langle S_i(t) \rangle\right]^2}}. \qquad (1)$$

A microstate of the system is delineated by the collective fluctuations of all grids, represented as an $N$-dimensional vector:

$$\delta S(t) = \begin{bmatrix} \delta S_1(t) \\ \delta S_2(t) \\ \mathrm{M} \\ \delta S_N(t) \end{bmatrix}. \qquad (2)$$

To construct the statistical ensemble of the complex system, we organize $T$ microstates into an $N \times T$ matrix $A$. The expression of the matrix element $A_{it}$ is $A_{it} = \delta S_i(t)/\sqrt{C_0}$, where $C_0 = \sum_{t=1}^{T}\sum_{i=1}^{N} \delta S_i^2(t)$ is the normalization factor. The sequence of columns in matrix $A$ corresponds to the evolution of these microstates. Further, the temporal correlation matrix $C$ and spatial correlation matrix $K$ can be defined by the matrix $A$

$$C = C_0 A^T \cdot A, \qquad (3)$$

$$K = C_0 A \cdot A^T, \qquad (4)$$

The correlation matrix $C$ has $T$ eigenvectors $V_J$ for $J = 1, 2, \ldots, T$., which can form an $T \times T$ unitary matrix $V = [V_1, V_2, \text{L}, V_T]$. The correlation matrix $K$ has $N$ eigenvectors $U_I$ for $I = 1, 2, \ldots, N$., which can form an $N \times N$ unitary matrix $U = [U_1, U_2, \text{L}, U_N]$. According to the singular value decomposition (SVD), the ensemble matrix $A$ can be factorized as

$$A = U \cdot \Sigma \cdot V^T, \tag{5}$$

where $\Sigma$ is an $N \times T$ diagonal matrix with elements $\Sigma_{IJ} = \begin{cases} \sigma_I, & \text{for } I = J \leq r, \\ 0, & \text{for otherwise,} \end{cases}$ and $r = \min(N, T)$. We can further rewrite the matrix $A$ as

$$A = \sum_{I=1}^{r} \sigma_I A_I^e = \sum_{I=1}^{r} \sigma_I U_I \otimes V_I, \tag{6}$$

where $A_I^e = U_I \otimes V_I$ is an $N \times T$ matrix with the element $\left(A_I^e\right)_{it} = U_{iI} V_{tI}$. We refer to the ensemble defined by $A_I^e$ as the eigen ensemble of the system. As $\sum_{I=1}^{r} \sigma_I^2 = 1$, $\sigma_I^2$ can be represented as the probability of the eigen ensemble $A_I^e$ in the statistical ensemble $A$. In conclusion, the statistical ensemble of system can be determined through a linear superposition of eigen ensemble $A_I^e$ with the probability of $\sigma_I^2$. From the statistical ensemble, we can obtain not only the eigen microstate $U_I$, but also its temporal evolution $V_I$.

**Text S2. Advanced weighted power spectral density method**

To elucidate frequency characteristics, Meng et al. (2023) further define an advanced weighted power spectral density $W_{PS}$ based on Welch's method (Welch, 1967):

$$W_{PS} = \int_f P(f) \times f \, df, \tag{7}$$

where $P(f)$ is the normalized spectral density and $f$ stands for the corresponding frequency derived from Fourier transform. Notably, higher $W_{PS}$ values indicate more rapid fluctuations, reflecting high-frequency changes.

**Text S3. Deseasonalization of data**

The deseasonalized data is calculated as

$$x_i^{'}(y,m) = x_i(y,m) - \frac{1}{N_y}\sum_{y} x_i(y,m), \tag{8}$$

where $x_i(y, m)$ stands for the data at the grid $i$ in the month $m$ ($m=1, 2, 3, …, 12$) of the year $y$ and $N_y$ is the number of years.

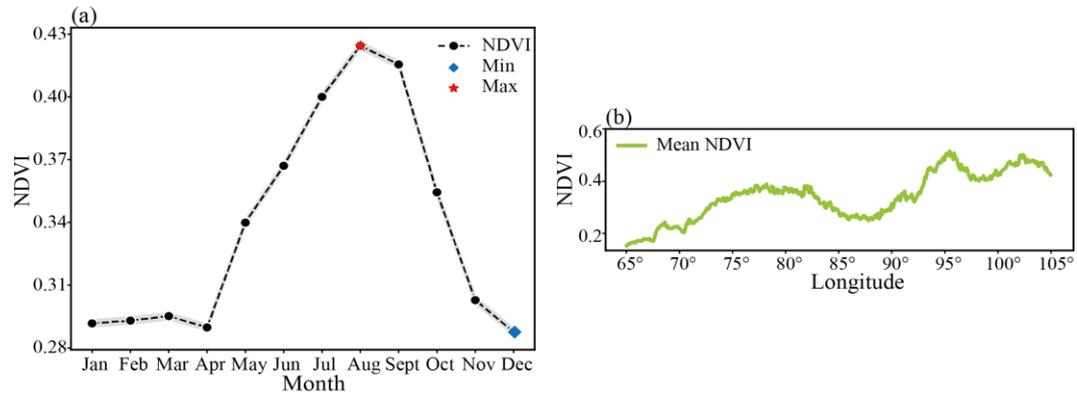

**Fig. S1. Spatiotemporal variability of NDVI.** (a) Monthly averaged NDVI with shading indicating a 95% confidence interval. (b) Average NDVI along longitude.

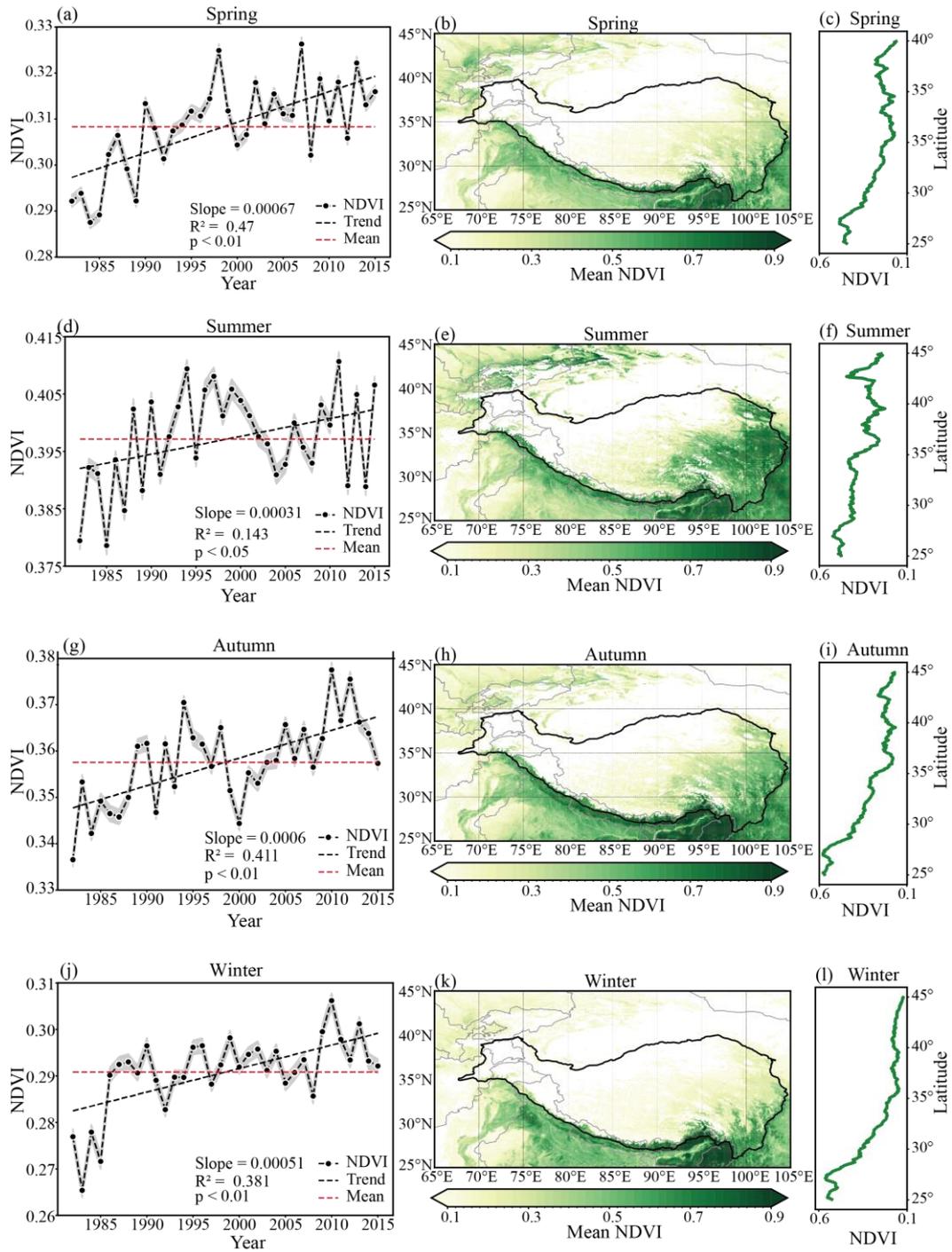

**Fig. S2. Seasonal variation of NDVI.** (a), (d), (g), (j) Temporal evolution of annual average NDVI across the four seasons: spring (March to May), summer (June to August), autumn (September to November), and winter (December to February). The shading indicates a 95% confidence interval. (b), (e), (h), (k) Spatial distribution of the mean NDVI across four seasons. (c), (f), (i), (l) Average NDVI along the latitude across the four seasons.

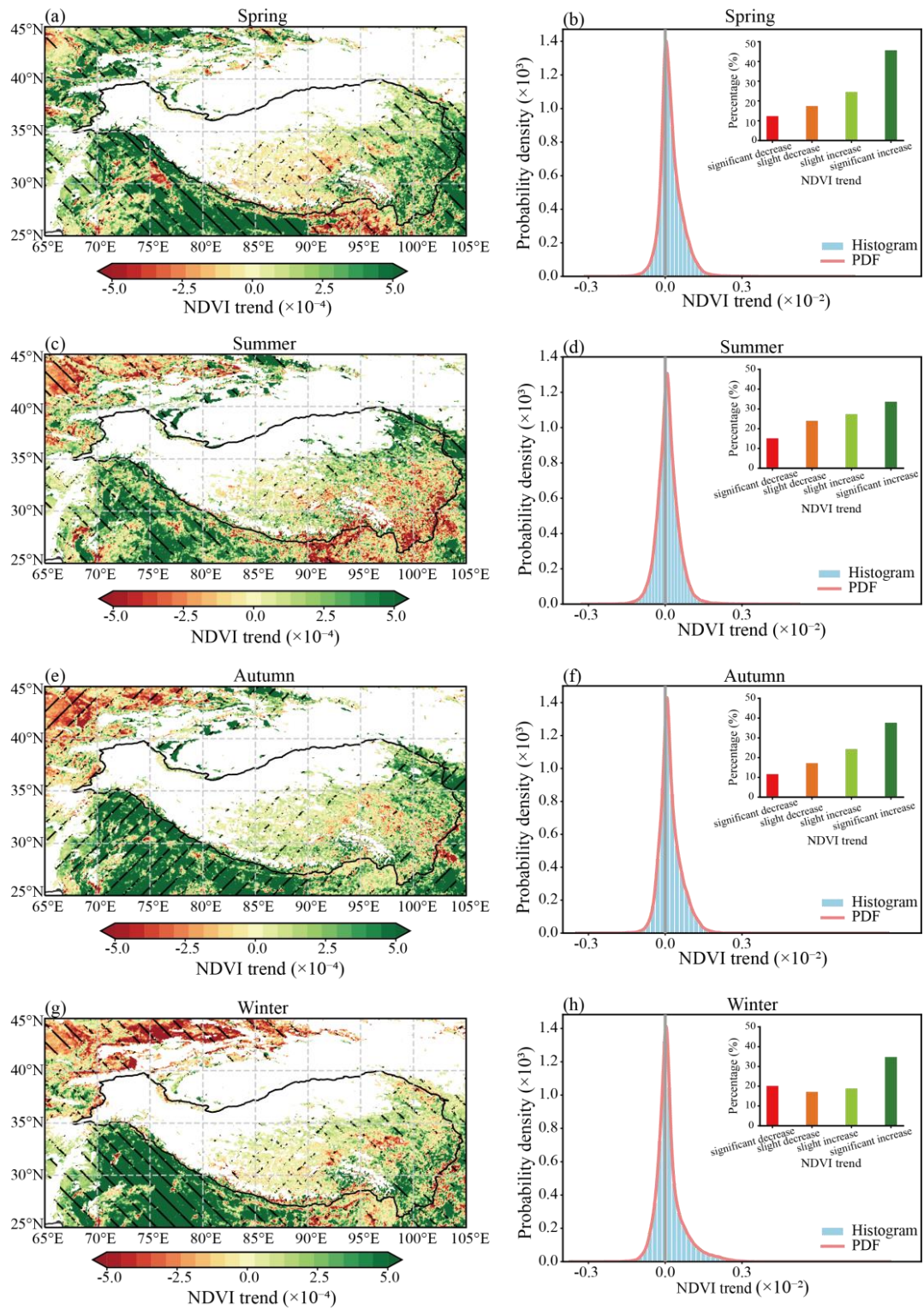

**Fig. S3. Trends of NDVI across four seasons from 1982 to 2015.** (a), (c), (e), (g) Spatial distribution of trends across the four seasons. (b), (d), (f), (h) Probability density distribution of trends across the four seasons. Shaded areas with diagonal lines indicate regions where the significance level is below 0.05.

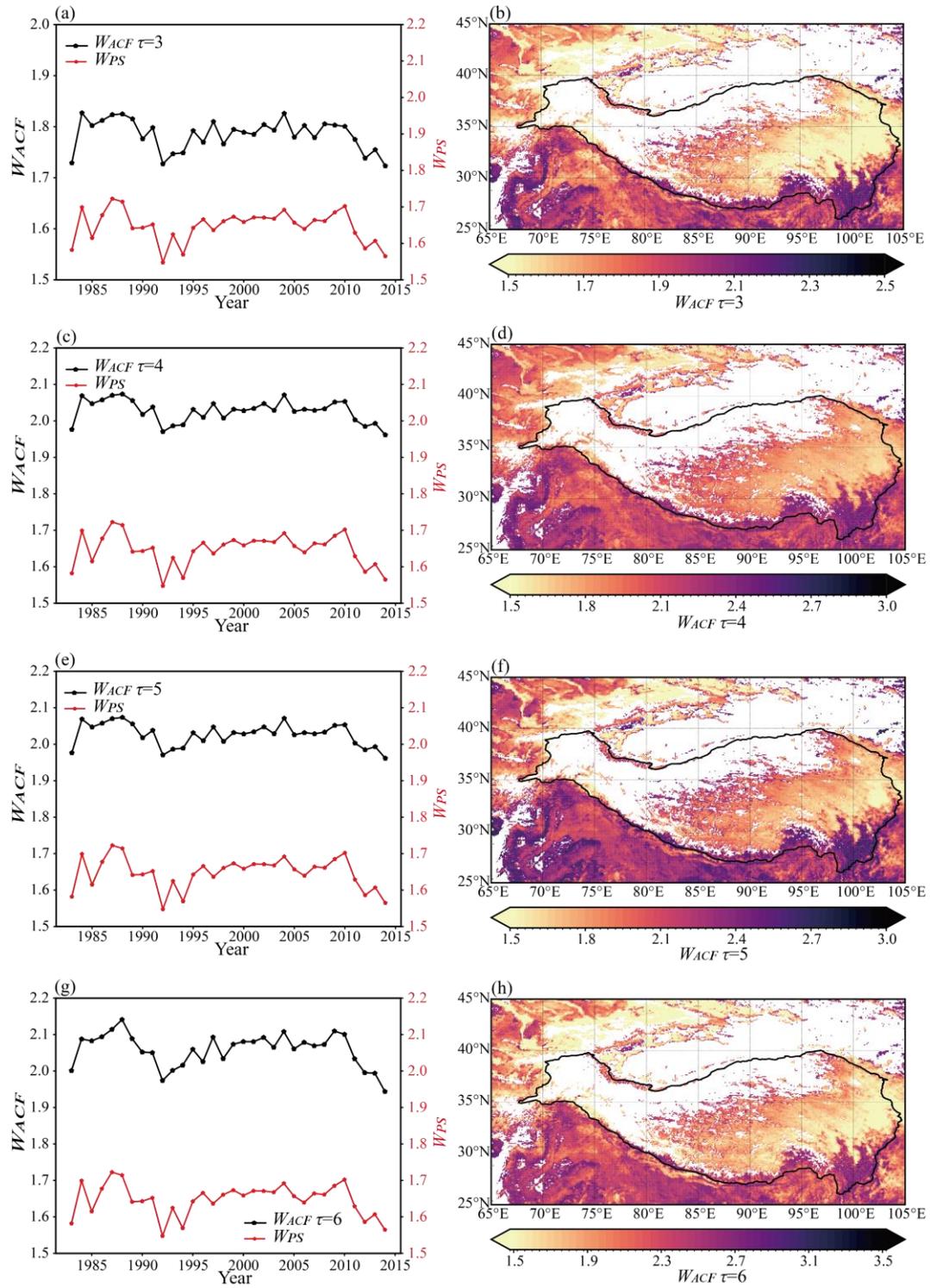

**Fig. S4. Temporal evolutions and spatial patterns of vegetation stability at different time lags $\tau$.** (a), (c), (e), (g) Temporal variations of the annual mean $W_{PS}$ and $W_{ACF}$ at different time lags $\tau$. (b), (d), (f), (h) Spatial patterns of $W_{ACF}$ at different time lags $\tau$.

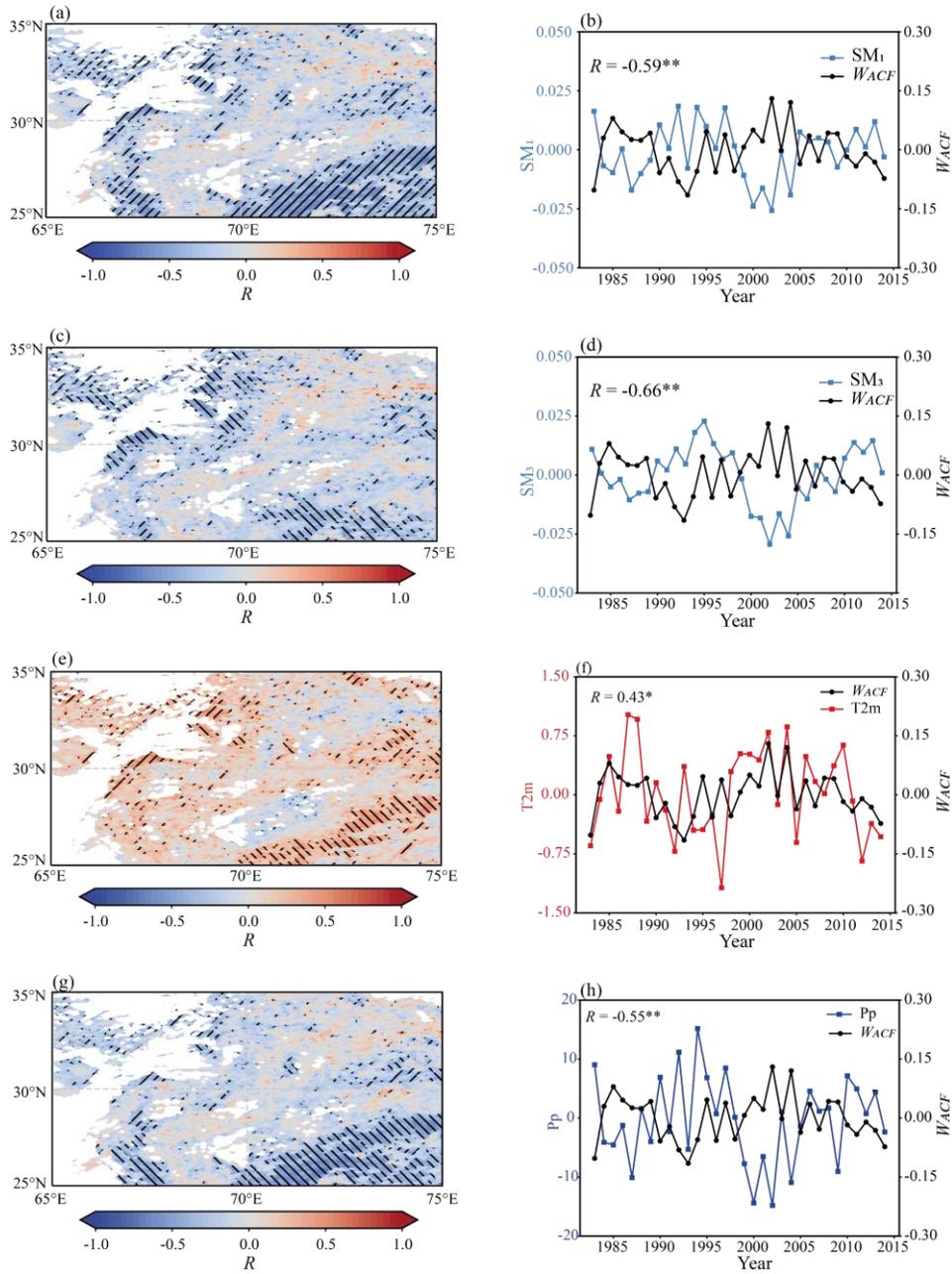

**Fig. S5. Analysis of the stability of vegetation variation.** (a) Spatial distribution of the correlation coefficient ($R$) between $W_{ACF}$ and the annual mean $SM_1$. (b) Temporal variations of the annual mean $SM_1$ and $W_{ACF}$ in the region spanning 25°N-35°N and 65°E-75°E, along with the correlation coefficient ($R$). (c), (e), (g) Same as (a), but for $SM_3$, T2m, and Pp, respectively. (d), (f), (h) Same as (d), but for $SM_3$, T2m, and Pp, respectively. Additionally, asterisks denote statistical significance: * $p<0.05$, ** $p<0.01$. Shaded areas with diagonal lines indicate regions where the significance level is below 0.05.

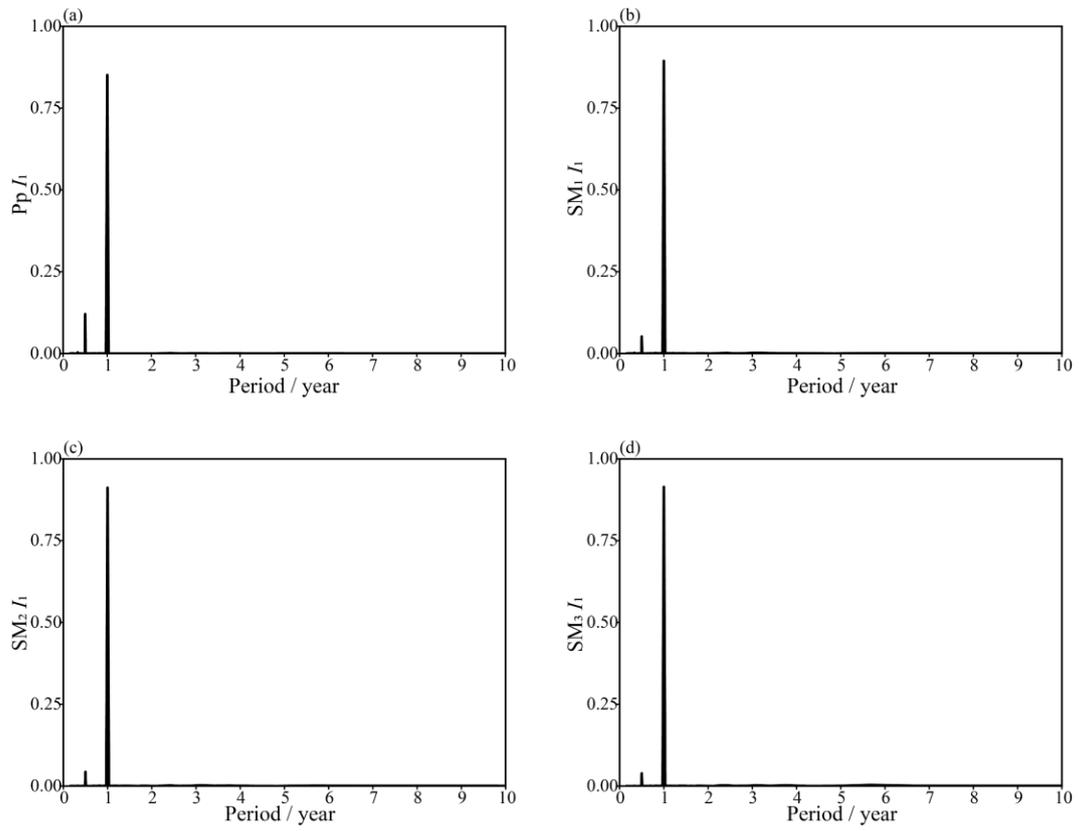

**Fig. S6. Power spectrum density of temporal evolutions of the first mode for Pp, SM$_1$, SM$_2$, and SM$_3$.**